\documentclass[namedreferences]{solarphysics}

\usepackage[hyperref,optionalrh,showbiblabels]{spr-sola-addons} 
\usepackage{graphicx}        

\usepackage{media9}
\usepackage{rotating}

\usepackage{amsmath}

\usepackage[dvipsnames]{xcolor}           

\chardef\us=`\_

\newcommand{\fract}[2]{\leavevmode\kern.1em
          \raise.5ex\hbox{\the\scriptfont0 #1}\kern-.1em
    \raise.15ex\hbox{\the\scriptfont0 /}\kern-.08em\lower.25ex\hbox{\the\scriptfont0 #2}}

\newcommand{\rd}{\mathrm{d}}

\newcommand{\rD}{\mathrm{D}}

\newcommand{\pderiv}[2]{\frac{\partial#1}{\partial#2}}

\newcommand{\deriv}[2]{\frac{\rd#1}{\rd#2}}

\newcommand{\Deriv}[2]{\frac{\rD#1}{\rD#2}}

\newcommand{\g}{\mathbf{g}}
\renewcommand{\k}{\mathbf{k}}

\newcommand{\bj}{\mathbf{j}}

\newcommand{\boldv}{{\mathbf{v}}}

\newcommand{\x}{\mathbf{x}}

\newcommand{\B}{{\mathbf{B}}}

\newcommand{\E}{{\mathbf{E}}}

\newcommand{\vdot}{{\boldsymbol{\cdot}}}
\newcommand{\vcross}{{\boldsymbol{\times}}}
\newcommand{\grad}{\mbox{\boldmath$\nabla$}}

\newcommand{\thth}{\hspace{1.5pt}}
\newcommand{\curl}{\grad\vcross}
\newcommand{\Curl}{\grad\vcross\thth}

\newcommand\Div{\grad\vdot\thth}

\newcommand{\kpar}{k_{\scriptscriptstyle\parallel}}

\newcommand{\bv}{Brunt-V\"ais\"al\"a}

  \renewcommand{\le}{\leqslant}
  \renewcommand{\ge}{\geqslant}

\begin{document}

\begin{article}
\begin{opening}

\title{Conversion and Smoothing of MHD Shocks in Atmospheres with Open and Closed Magnetic Field and Neutral Points}

\author[addressref={aff1},corref,email={jamon.pennicott@monash.edu}]{\inits{J.~D.}\fnm{Jamon D.}~\lnm{Pennicott}}
\author[addressref=aff1,email={paul.cally@monash.edu}]{\inits{P.~S.}\fnm{Paul S.}~\lnm{Cally}}

\address[id=aff1]{School of Mathematics, Monash University, Clayton, Victoria, Australia}

\runningauthor{Pennicott and Cally}
\runningtitle{Conversion and Smoothing of MHD Shocks}

\begin{abstract}
Planar acoustically dominated magneto\-hydro\-dynamic waves are initiated at the high-$\beta$ base of a simulated 2D isothermal stratified atmosphere with potential magnetic field exhibiting both open and closed field regions as well as neutral points. They shock on their way upward toward the Alfv\'en-acoustic equipartition surface $a=c$, where $a$ and $c$ are the Alfv\'en and sound speeds respectively. Expanding on recent 1.5D findings that such shocks mode-convert to fast shocks and slow smoothed waves on passing through $a=c$, we explore the implications for these more complex magnetic geometries. It is found that the 1.5D behaviour carries over to the more complex case, with the fast shocks strongly attracted to neutral points, which are disrupted producing extensive fine structure. It is also observed that shocks moving in the opposite direction, from $a>c$ to $a<c$, split into fast and slow components too, and that again it is the slow component that is smoothed.
\end{abstract}
\keywords{Waves, Magneto\-hydro\-dynamic; Waves, Shock; Heating, Chromospheric}
\end{opening}

\section{Introduction}
     \label{S-Introduction} 

\subsection{Linear mode conversion} \label{SS-linIntro}
It is well-established that linear fast and slow magneto\-hydro\-dynamic (MHD) waves may partially mode-convert between each other on Alfv\'en-acoustic equi\-partition surfaces $a=c$, where $a$ is the Alfv\'en speed and $c$ is the sound speed \citep{SchCal06aa}. The result is based on a sophisticated ray theory incorporating WKB connection coefficients \citep{TraKauBri03aa,TraBriRic14aa}, and is confirmed by exact solution \citep{HanCal09aa} and simulations \citep[e.g.,][]{NutSteRot10aa,RijRajPrz16aa,RieVanCal19aa}. This solves the sunspot p-mode absorption problem \citep{BraDuvLab88aa,CalCroBra03aa}, where $a=c$ is typically sub-photospheric, with implications for wave-coupling of the solar interior and atmosphere \citep{CalMor13aa}. Away from sunspots, equipartition surfaces normally reside in the chromosphere, possibly taking complex shapes, giving rise to a rich array of wave behaviours \citep{NutSteRot12aa,KhoCal13aa}.


\subsection{Mode conversion of shocks} \label{SS-shockIntro}
Several recent articles have addressed how fast/slow mode conversion carries over to shock waves. Invoking WKB-based arguments, \cite{Nun19aa} predicted analytically that shocks would mode-convert at $a=c$ in the same way that linear waves do, but that both emerging fast and slow MHD waves would be smoothed, i.e., {they are} no longer shocks. 

Simulations in 1.5D by \cite{PenCal19aa} however revealed that in fact the emerging fast wave remains a shock, though the slow wave is smoothed (for non-zero attack angle between the shock normal and the magnetic field). A major reason for the discrepancy is that the shock incident on the equipartition layer is seen to drag it along for an extended period (some seconds), invalidating an assumption of N\'u\~nez's. Subsequently, \cite{SnoHil20vq} confirmed these findings, and extended them to two-fluid shocks in a partially-ionized plasma, discovering several further behaviours dependent on collision strengths.

Shocks in an initially uniform MHD plasma are normally described in terms of the compression ratio $X=\rho_2/\rho_1$, where the subscript `1' denotes the pre-shocked plasma and `2' the post-shocked. Standard theory was developed by \cite{BazEri59aa} and has been summarized in any number of text books \citep[e.g.,][]{Pri82aa} since. Just as for hydrodynamic shocks, it turns out that $1<X<(\gamma+1)/(\gamma-1)$, i.e., $1<X<4$ for $\gamma=5/3$. By conservation of mass, $v_{x2}=v_{x1}/X$, where $x$ is the direction perpendicular to the shock front, so the normal velocity slows down relative to the incident velocity by no more than this limiting factor. 

The three shock types -- fast, intermediate (Alfv\'en) and slow -- reduce to those three linear MHD wave types in the limit $X\to1^+$. However, as the shock strength $X>1$ increases, the slow and intermediate shocks annihilate at some $X_{\mbox{crit}}$ above which only the fast wave persists all the way up to $X=(\gamma+1)/(\gamma-1)$, corresponding to infinite Mach number. \cite{PenCal19aa} use this as a partial explanation for the disappearance of the slow shock but persistence of the fast shock in their 1.5D simulations as the shock crosses $a=c$. In the following sections, we explore this further in a complex magnetic geometry, with particular attention paid to the open and closed field regions and the neighbourhood of neutral point. 

The purpose of this article is to explore the implications of these findings for shock waves in more complex magnetic geometries, as typically are found in the Sun's inner and outer atmospheres. For simplicity, the models are two-dimensional (2D), which is sufficient for our purposes. The introduction of a third dimension would add fast-Alfv\'en coupling as well \citep{CalHan11aa,KhoCal12aa}, which would complicate the results and in any case is beyond our scope here. Only the one-fluid ideal MHD case will be considered.

Specifically, we investigate models with a mixture of open and closed magnetic field that exhibit X-type neutral points. The contrasting behaviours in open and closed field and the focusing, defocusing and mode conversion of fast and slow waves near the neutral point will be explored. Particular attention will be paid to the effect of attack angle, the complex dragging of the $a=c$ layer in the heterogeneous model, and the role of a secondary $a=c$ ``island'' around the neutral points. Local shock and electrical heating is addressed. We will compare and contrast linear and shock wave behaviours, especially their mode conversion properties. The smoothing or otherwise of shocks crossing the equipartition layers will be of special interest.

\section{Model}
      \label{S-model}

In a prominent recent review of the solar chromosphere, \cite{CarDe-Han19aa} state that \emph{``Magnetoacoustic slow-mode shock waves permeate the chromosphere, both in the internetwork quiet Sun and in strong magnetic field regions like network and plage. They play a key role in the momentum and energy balance of the chromosphere\ldots''}, and highlight the impact of small-scale magnetic fields as an emerging issue of relevance to the energy balance. The disparity of scales between network and small magnetic dipoles in particular results in a mix of magnetic field lines that are open and closed within the chromosphere. This inspires us to explore the contrasting behaviours of shocks in open and closed field and around associated neutral points, with particular attention paid to mode conversion, shock smoothing, and development of fine structure that may enhance local heating.

\subsection{Equilibrium Model} 
  \label{S-eqbm}

In terms of horizontal and vertical coordinates $(x,z)$, we express a magnetic field $\B$ in terms of its flux function $A(x,z)$,
\begin{equation}
    B_x=\pderiv{A}{z}, \qquad B_z=-\pderiv{A}{x}.  \label{B}
\end{equation}
For potential field, $A$ is harmonic, $\nabla^2A=0$.
 A suitable separable choice is the well-known $L$-periodic potential magnetic arcade \citep{Pri82aa} supplemented with a uniform inclined field,
\begin{equation}
    A(x,z) = B_u (z \sin\theta-x\cos\theta)-k^{-1}B_0 \cos(k x) e^{-k z}     \label{A}
\end{equation}
where $k=2\pi/L$. 
The field strength parameters $B_0$ and $B_u$ represent respectively a pure arcade and a superimposed uniform field that is inclined at angle $\theta$ to the vertical. At the base, the vertical field component is $B_z(x,0)=B_u\cos\theta+B_0\sin k x$, and $\B\to B_u(\sin\theta,\cos\theta)$ as $z\to\infty$. The inclined field gives us some control over attack angle.

An isothermal gravitationally stratified atmosphere with sound speed $c$ and pressure $p_0\, e^{-z/H}$ is sufficient to illustrate the relevant phenomena. The scale height $H=c^2/\gamma g$, where $\gamma=5/3$ is the adiabatic index and $g=274$ $\rm m\,s^{-2}$ is the gravitational acceleration. For the pure arcade model, $B_u=0$, Alfv\'en speed increases with height provided $2kH<1$, i.e., $4\pi H<L$, which we shall generally enforce.

\begin{figure}[thbp]
    \centering
     \includegraphics[width=.75\textwidth]{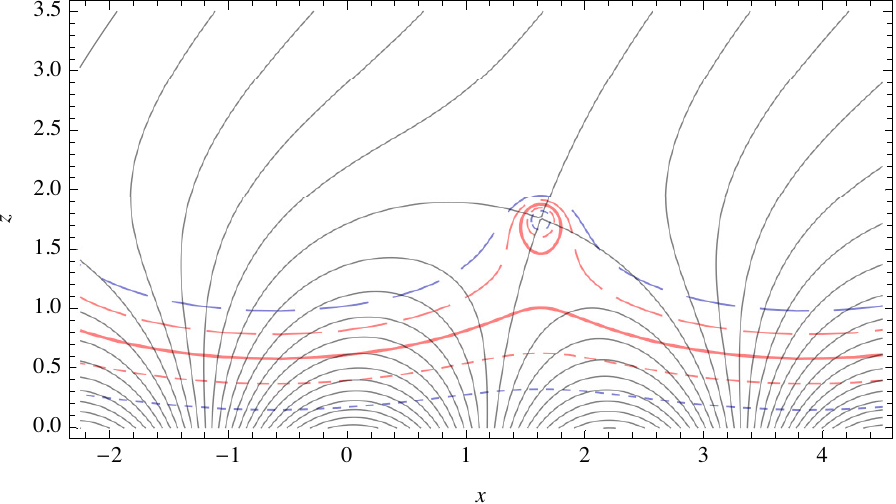}\\[6pt]
    \includegraphics[width=.75\textwidth]{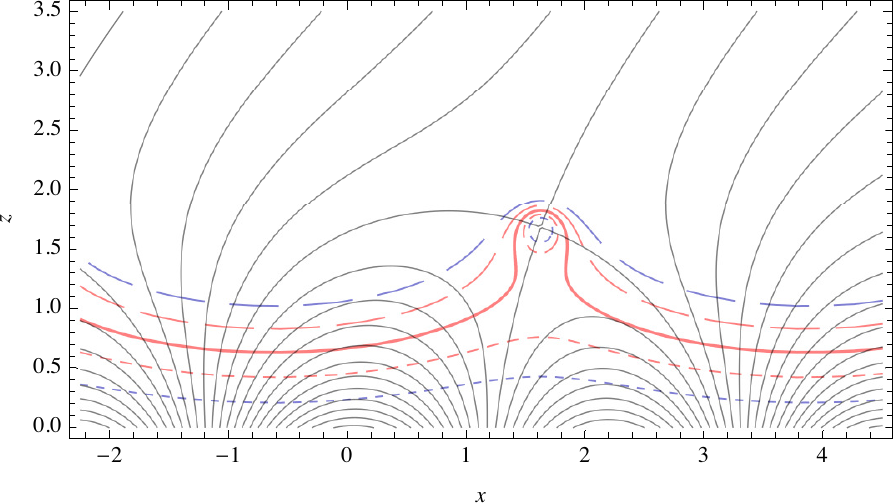}
    \caption{Two representative magnetic and thermal scenarios, corresponding to the case $L=4.5$~Mm, $c=8.96$~$\rm km\,s^{-1}$, 
    $B_0=520$~G (top; ``island'') or 470~G (bottom; ``promontory''),
    $B_u=45$~G with $\theta=40^\circ$, $\gamma=5/3$, and base pressure $p_0=10$~kPa. With these values, the density scale height is $H=176$ km and $2kH=0.49<1$. The magnetic field lines are in black, and the X-point is evident, as are the separate open and closed field regions. The full red curve is the $a=c$ layer. There is a secondary equipartition island around the neutral point in the upper panel ($B_0=520$~G), which connects as a promontory in the lower panel (470~G), because weaker magnetic field moves the $a=c$ layer upward, or outward from the neutral point. The red long-dashed curve is $a/c=\sqrt{2}$, the blue long-dashed curve is $a/c=2$, the red short-dashed curve is $a/c=1/\sqrt{2}$, and the blue short-dashed curve is $a/c=1/2$. The model has period 4.5 Mm in $x$, with 1.5 periods shown.}
    \label{fig:Xexample}
\end{figure}

Figure \ref{fig:Xexample} illustrates two typical scenarios (``island'' and ``promontory'') on a more-or-less chromospheric length scale, notwithstanding the lack of a transition region, with typical chromospheric sound speed, pressure, and magnetic field strength (as detailed in the figure caption). The distinction between the island and promontory cases in this instance hinges on a small difference in field strength; stronger field lowers the $a=c$ layer and causes the island to decouple from it. This is not an important distinction. Simulations with either scenario produce similar results. For this reason, we shall address the generic class of models illustrated by the upper panel of Fig.~\ref{fig:Xexample}, consisting of a roughly level equipartition layer with an isolated neutral point and equipartition island above.

We shall see that a sufficiently strong incident shock wave significantly modifies the location of the equipartition surfaces, and can turn an island into a promontory. 

If the sound speed were increased by a factor of 10 say, the density scale height $H=c^2/\gamma g$ (and temperature) would increase by a factor of 100, so if $L$ were also increased by factor 100, the scaling would be homologous. Moreover, if the base plasma pressure $p_0$ and the magnetic pressure $B^2/2\mu$ were also scaled in lock-step, the position of the $a=c$ layer would be identical in this rescaled model. Such a high temperature large length-scale scenario would be more coronal in character, but will not be considered explicitly.

\subsection{Equations} 
  \label{S-eqns}
The evolution of the system is described in terms of the density $\rho$, fluid velocity $\boldv$, plasma pressure $p$, specific internal energy density $\epsilon=p/((\gamma-1)\rho)$, gravitational acceleration $\g$, magnetic field $\B$, electric field $\E=-\boldv\vcross\B 
$, current density $\bj=\mu^{-1}\Curl\B$ and sound speed $c=\sqrt{\gamma p/\rho}$ with adiabatic index $\gamma=5/3$. The equation of state is taken to be that of a perfect fully ionized hydrogen gas for simplicity. 

These quantities are governed by the standard single-fluid two-dimensional (2D) non-linear ideal magneto\-hydro\-dynamic (MHD) equations under gravity in the negative $z$ direction. Specifically, these are in effect the familiar continuity, momentum, induction and internal energy equations
\begin{subequations}
\begin{gather}
    \Deriv{\rho}{t}+\rho \Div\boldv=0,\\
    \rho \Deriv{\boldv}{t}=-\grad p+\rho\g+\bj\vcross\B,\\
    \pderiv{\B}{t}=-\curl\E,\\
    \rho\Deriv{\epsilon}{t}=-p\Div\boldv 
\end{gather}
\end{subequations}
which are solved subject to periodic boundary conditions in the horizontal direction $x$. The top and bottom boundary conditions are described below. The initial state is isothermal ($c=\text{constant}$) stratified equilibrium permeated by the potential magnetic field described above. These MHD equations are linearized only for the ray calculations of Sec.~\ref{D-X}.

\subsection{Numerical Scheme} 
  \label{S-numerics}
  
The numerical simulations are performed using the Lare2d code \citep{ArbLonGer01aa} to solve the single fluid nonlinear MHD equations. The numerical grid contains 512 cells in the horizontal direction and 1024 cells in the vertical direction, and the code's default shock viscosities were used.  The grid is uniform in the $x$ direction and spans from $-2.25$ Mm to 2.25 Mm, giving a resolution of $\Delta x = 8.79$ km, corresponding to about 45 points across the neutral point island for example.  The grid is stretched in the $z$ direction and spans from 0.00 Mm to 36.0 Mm, though only the bottom 3.5 Mm will be displayed. The stretching function acts on $\Delta z$, such that
\begin{equation}
    \Delta z(z) =   \left(1 + f\left(1 + \tanh\frac{z - L_z}{W}\right)\right)\,\delta z
\end{equation}
where $\delta z = 5.86$ km, $f = 10.0$, $L_{z} = 4.50$ Mm, and $W = 0.60$ Mm. This gives vertical spacings $\Delta z$ ranging from 5.86 km at $z=0$ through 5.89 km at $z=2$ Mm to 9.9 km at $z = 3.5$ Mm, which is the domain of interest.

The horizontal boundary conditions are periodic, whilst the vertical boundaries are closed.  A velocity damping profile is implemented for $z \ge 4.5$ Mm, which suppresses the outgoing waves and prevents artificial reflections from the top of the computational box. In effect, velocity update above this level was supplemented by a damping term $-\gamma_{\rm d}\boldv$ where the damping factor $\gamma_{\rm d}$ linearly increases with height from zero at 4.5 Mm, with no apparent ill-effects on the lower levels. No artificial cooling or radiative losses are employed throughout the atmosphere.

The model adopted is an isothermal gravitationally stratified 2D atmosphere with an overlying potential magnetic field structure.  The magnetic field is described by Equations (\ref{B}) and (\ref{A}) with $B_{0} = 520$ G, $B_{u} = 45$ G and $\theta = 40^{\circ}$. The gravitational acceleration $\mathbf{g} = (0,-g)$ is defined by the piecewise function
\begin{equation}
  g =
  \begin{cases}
  g_{0} & z \le z_{1} \\
  g_{0}(1 + \cos{(\pi \frac{z-z_{1}}{z_{2}-z_{1}})}) & z_{1} < z \le z_{2} \\
  0     & z > z_{2}
  \end{cases}
\end{equation}
 where $g_{0} = 274$ m s$^{-2}$, $z_{1} = 3.0$ Mm, and $z_{2} = 4.5$ Mm. 
 
 The uniform sound speed is $c = 8.96$ km s$^{-1}$, the scale height $H = 176$ km and the acoustic cut-off frequency $\omega_{c} = c/(2H) = 4.0$ mHz.  The initial $a=c$ contours produced by this atmosphere can be seen in Figure \ref{fig:Xexample}, which shows 1.5 periods in $x$. The extended top of the box with zero gravity allows waves to escape and to not reflect back into the region of interest.
  
To drive the waves, a purely vertical velocity pulse is implemented for half a period along the bottom boundary of the form
\begin{equation}
    v_{z}(x,0) = v_{0} \sin{(\omega t)},
    \label{driver}
\end{equation}
\citep[as in][]{PenCal19aa}, where $v_{z}$ is vertical velocity, $v_{0}$ is amplitude, $\omega/2\pi = 10.0$ mHz is the frequency, and $t$ is time.  Due to the relatively short driver period, this corresponds to a broad frequency spectrum centered around zero.  However, the acoustic cut-off will only allow waves above $\omega_c/2\pi = 4.0$ mHz to propagate upwards.
  
\section{Waves Near Neutral Points}\label{S-X} 
Magneto\-hydro\-dynamic waves near magnetic neutral points have been studied in detail over many years, inspired by space-based observations of the ubiquity of MHD oscillations in the solar corona and the need to understand the creation of small length scales so that they can dissipate their energy and heat the atmosphere. 

\cite{McLHoo04aa} found that the linearized fast wave ``wraps itself around the null point due to refraction'' in a 2D zero-$\beta$ plasma, producing a spike in current density and probable local heating. \cite{NakMel09aa} and \cite{VanKupYua16aa} noted a possible link to Quasi-Periodic Pulsations (QPP) associated with flaring energy release in loop systems, where X-type neutral points are a feature of standard models.

\cite{McLHoo06aa} included a finite sound speed and discovered that fast/slow mode conversion takes place on the equipartition surface surrounding the null point (erroneously identified as the $\beta=1$ surface rather than $a=c$, but the distinction is minor since $\beta=2/\gamma=1.2$ at $a=c$). They also found that some of the generated slow wave was able subsequently to leave the region along separatrices. \cite{McLHoode-11lb} review the state of knowledge \textit{c.}~2010. More recently, 3D simulations by \cite{TarLin19aa} confirmed these general behaviours in a coronal context, but suggested that associated shock and ohmic heating is one-to-two orders of magnitude below what is required to supply radiative losses from coronal bright points.

Using conformal mapping techniques, \cite{Nun17aa} analytically investigated the propagation of fast waves around a potential-field neutral point in the low-$\beta$ limit, elucidating the particular mathematical properties of rays, which describe logarithmic spirals, and wavefronts that exhibit more complex behaviour. They form shocks from linear waves in a finite time if the wavefront is convex, due to the appearance of caustics. This adds mathematical detail to the simulations of \cite{McLDe-Hoo09aa}. Both confirm that nonlinearities bring new features to the model. 

However, N\'u\~nez's calculations say nothing about mode conversion, as there is no equipartition surface, and therefore they do not apply inside the $a=c$ contour where a fraction of the wave will be slow and therefore not bound to the neutral point. Because of the X-type structure of the magnetic field in this region, the attack angle of a spiralling ray to the field lines at $a=c$ will take on very different values around the circumference, leading to a complex conversion picture, so numerical simulation is probably the only feasible solution strategy.


Most previous studies concentrate on the coronal context. However, similar processes on a much smaller scale are to be expected in the chromosphere as well, as discussed earlier.

Our main focus is the behaviour of waves and shocks as they pass through and sometimes advect the equipartition surfaces, rather than on the neutral points themselves. We therefore refrain from introducing additional physics more applicable to that topic, such as reconnection, although the code can address such non-ideal phenomena too. For example, in the Hall regime, X-type neutral points subject to impinging fast waves are found using Lare2d to exhibit multiple additional transient nulls and hence enhanced reconnection rates \citep{ThrParDe-12aa}, which is beyond our scope. 

Although linear and nonlinear waves near neutral points have been extensively examined previously, we explore them further here in the broader context of open and closed field, mode conversion and shock smoothing at $a=c$, equipartition layer dragging, and implications for heating mechanisms and locations.

\section{Results} 
      \label{S-results} 
      
\subsection{Linear Case} 
  \label{S-linear-sim}
For comparison, we first presents results for the linear wave case, so that the consequences of nonlinearity in subsequent simulations are clear.

A slab-like driver is implemented along the bottom boundary as described by Equation (\ref{driver}) with amplitude $v_{0} = 5.0$ m s$^{-1}$.  This will ensure wave amplitudes remain in the linear regime throughout the simulation in the area of interest.  Due to the high $\beta$ level at the base of the computational box, the majority of the driver's energy is imparted into the fast (acoustic) wave, and negligible energy is directed into the slow (magnetic) wave.

\begin{figure}[htbp]
    \centering
    \includegraphics[width=\textwidth]{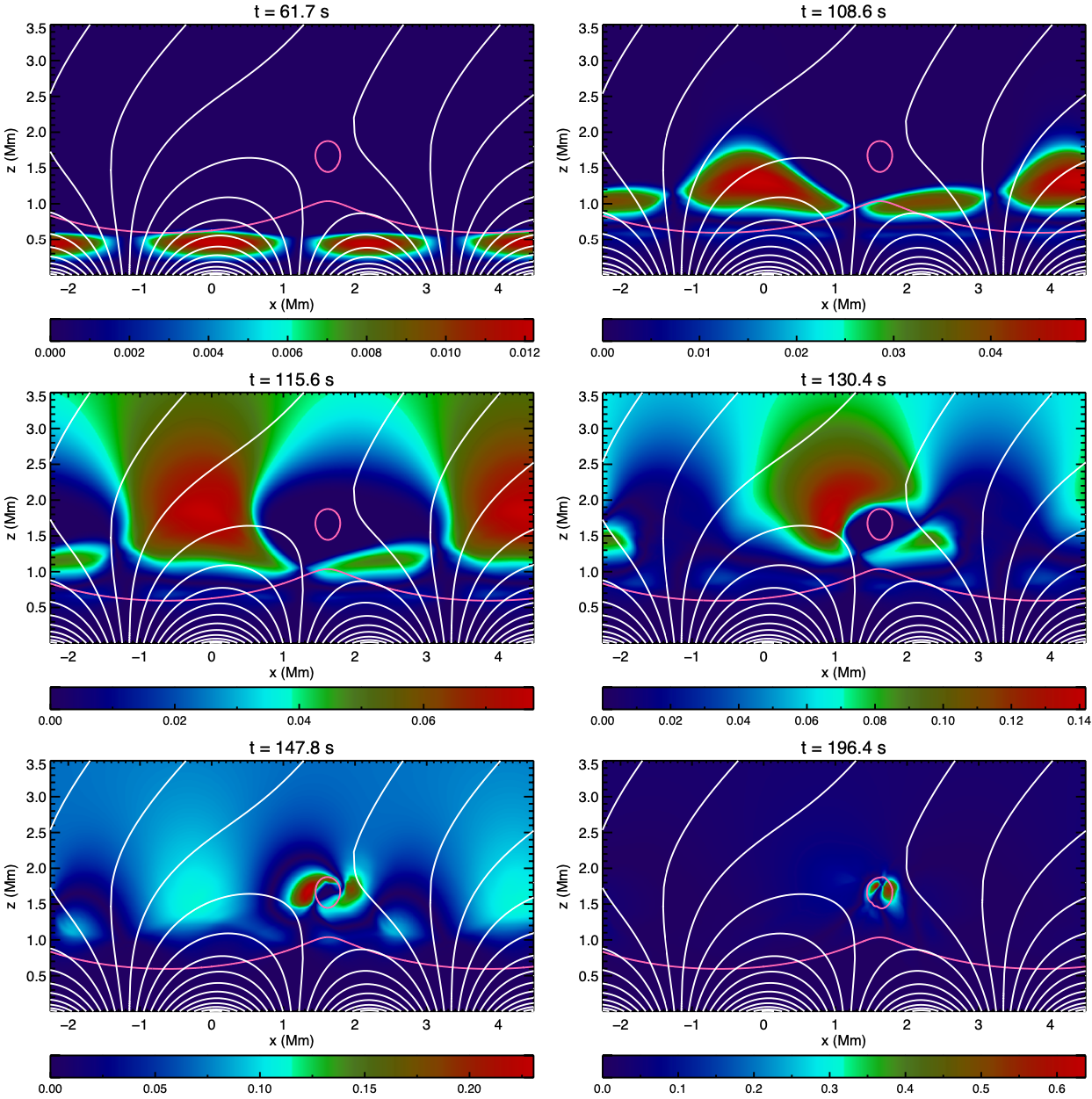}
    \caption{Timeline of $v_{\perp}$ (absolute value) snapshots for the linear case measured in km s$^{-1}$ with corresponding time in seconds displayed in each panel's title.  The superimposed white lines depict various magnetic field lines and the $a=c$ contours are represented in pink. Note that initially there is a roughly horizontal equipartition layer, and a disjoint island. With other field parameters that island may be a promontory. We notice the following evolution: $t=61.7$ s, the driven fast pulse has not yet reached the equipartition level; $t=106.6$ s, it has passed through the equipartition and is starting to refract towards the low-field region centred on the neutral point; $t=115.6$--$147.8$ s, the refraction progresses; $t=196.4$ s, the $v_\perp$ response is highly localized inside the neutral point island.}
    \label{fig:vperp_linear}
\end{figure}

\begin{figure}[htbp]
    \centering
    \includegraphics[width=\textwidth]{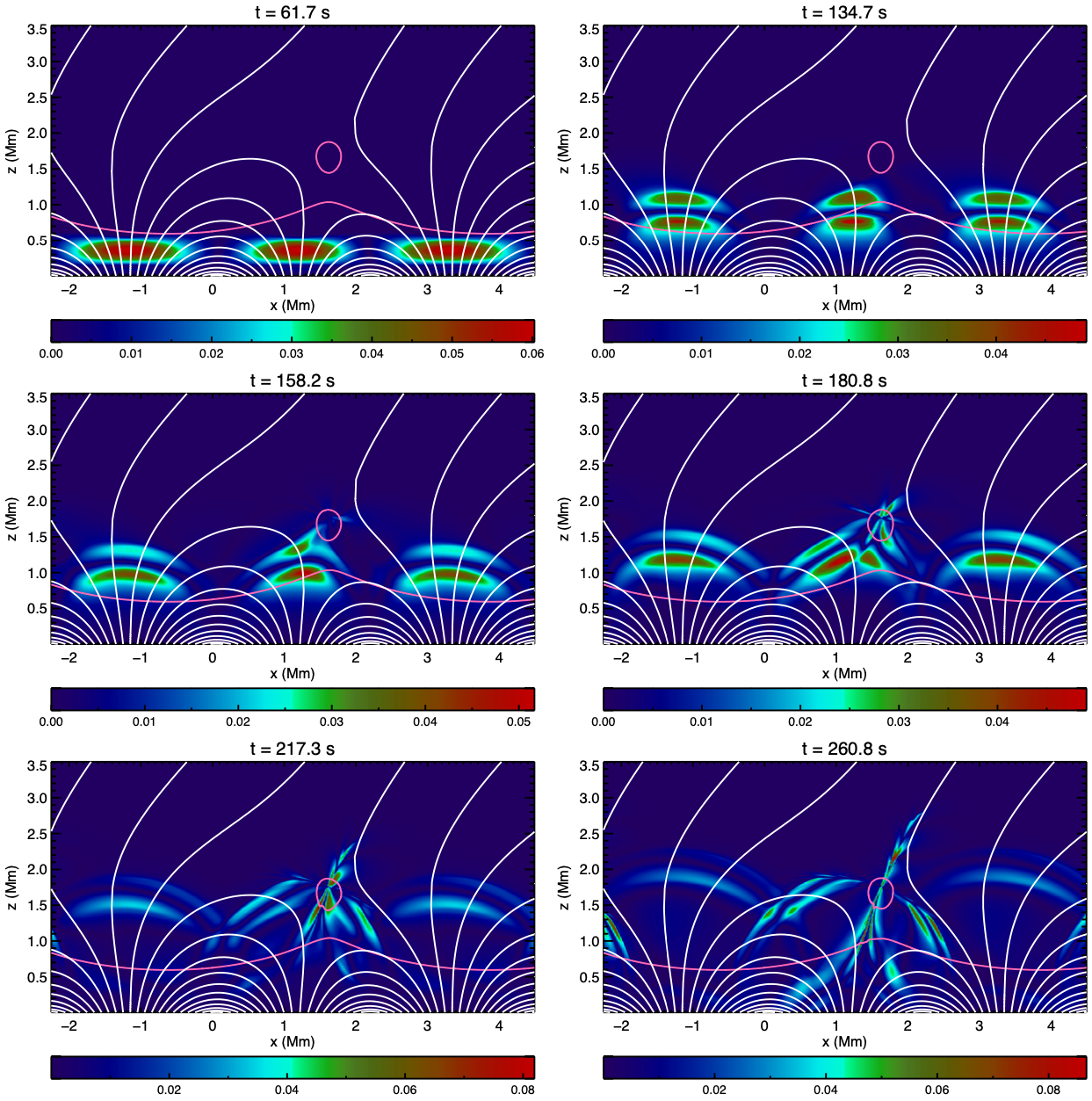}
    \caption{Timeline of $v_{\parallel}\sqrt{\rho}$ (absolute value) snapshots for the linear case measured in kg$^{1/2}$ m$^{-1/2}$ s$^{-1}$ with corresponding time in seconds displayed in each panel's title.  The superimposed white lines depict various magnetic field lines and the $a=c$ contours are represented in pink. We notice the following evolution: $t=61.7$ s, the driven fast pulse has not yet reached the equipartition level; $t=134.7$ s, the pulse has passed through $a=c$ and the plotted $v_\parallel$ signal is clearly associated with the slow wave, based on its slow progress and small vertical wavelength; $t=158.2$--$260.8$ s, the slow wave propagates upward uneventfully in the open field but progressively more field-aligned velocity structure is generated at the neutral point and aligns with the separatrices (see Fig.~\ref{fig:Xexample}). }
    \label{fig:vpar_linear}
\end{figure}

A timeline of events is produced in Figures \ref{fig:vperp_linear} and \ref{fig:vpar_linear}, which depict both the perpendicular and parallel (to the magnetic field) plasma velocities $v_\perp$ and $v_\parallel$. The parallel velocities are scaled by $\sqrt{\rho}$ in order to more easily view the acoustic waves present in all areas of the domain.  These plots are the primary signatures used to determine the presence of fast/slow and acoustic/magnetic MHD waves.

In the low-$\beta$ limit, $a\gg c$, fast waves are entirely transverse to the magnetic field and slow waves are polarized parallel to it, so the $v_\perp$/$v_\parallel$ split is ideal for distinguishing between them. However, this is not so where $a\lesssim c$. Near $a=c$ in particular, it does not even make sense to try to decouple fast and slow modes, as this is where they are physically coupled and exchanging energy. Nevertheless, the $v_\perp$/$v_\parallel$ split is useful in such figures as it allows us to unambiguously distinguish the waves high above the equipartition level and identify the fates of the separate modes. Associated animations elucidate this further by allowing us to easily see separate waves moving upwards along open field lines in the upper regions (near-parallel slow waves) and reflecting back downwards (near-transverse fast waves) elsewhere. 

Note that the times displayed in Figs.~\ref{fig:vperp_linear} and \ref{fig:vpar_linear} are deliberately chosen to not correspond, since the slow and fast evolutions that they respectively illustrate naturally move at different rates.

The rightmost boundary of the computational box sits at $x = 2.25$ Mm, however the plots displayed in Figs.~\ref{fig:vperp_linear} and \ref{fig:vpar_linear} stretch to $x = 4.5$ Mm. This enlarged window allows us to view the range of behaviours across the domain more clearly.

\begin{sidewaysfigure}
    \centering
    \includegraphics[width=\textwidth]{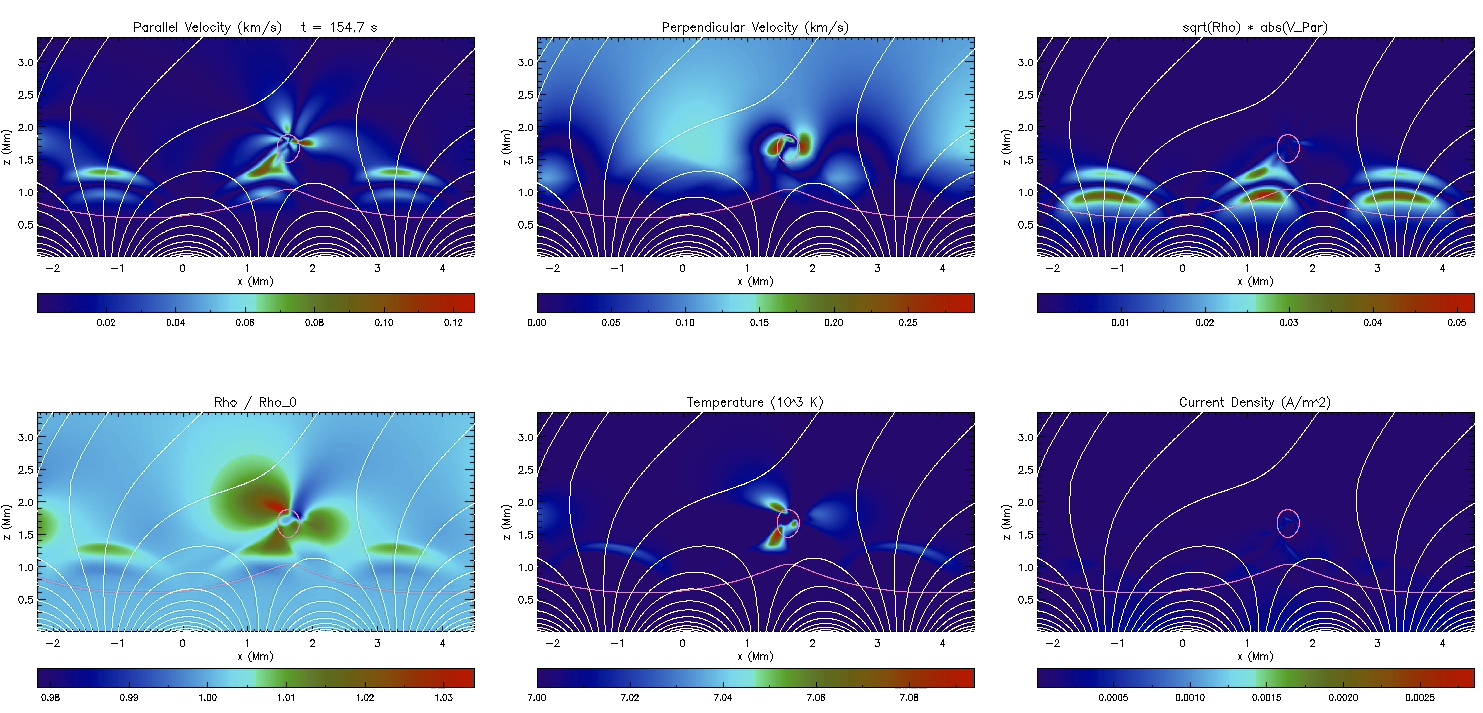}
    \caption{Snapshot at $t=154.7$ s from an animation (linear.mp4) of the linear case, showing respectively parallel velocity $v_\parallel$, perpendicular velocity $v_\perp$, scaled perpendicular velocity  $\sqrt{\rho}\,v_\perp$, density relative to the background state $\rho/\rho_0(z)$, temperature $T$ (in kK) and $y$ component of the current density, $j_y$. All quantities are shown as absolute values. The full animation is available as Electronic Supplementary Material.}
    \label{fig:lin_anim}
\end{sidewaysfigure}

Figure \ref{fig:lin_anim} is a frame from an animation of the linear case that shows both parallel and perpendicular velocity components as well as density, temperature and current density. The separate slow waves propagating steadily upward in the open field are clearly very distinct from the fast waves turning over and converging on the neutral point.

Shortly after the driver activates, the initially flat wavefront it produces begins to experience variation in its horizontal structure. Areas with higher magnetic field strength and consequently higher Alfv\'en velocities cause the fast wave to propagate faster. Various parts of the wavefront therefore reach the equipartition layer asynchronously. This is further influenced by the initial variation in height of the equipartition layer.

Eventually, all parts of the wavefront reach the equipartition layer and all are engaged in mode conversion/transmission. In areas of closed magnetic field, the generally large attack angle between the wave vector and the magnetic field sees significant conversion observed, where the majority of the energy in the acoustic (fast) wave is transmitted into the magnetic (fast) wave above $a=c$.  In contrast, areas of open magnetic field correspond to smaller attack angles and significant acoustic-to-acoustic (fast-to-slow) transmission is observed.

Following this process, the fast wave around the central axis begins to propagate fastest across the field and a section quickly veers towards the neutral point (seen clearly in the perpendicular velocity at about $t=115$ s in the linear wave animation). This is the normal wave propagation effect of refraction away from regions of higher wave speed. Soon after, the fast wave turns over as it refracts away from the ever increasing Alfv\'en velocity high in the atmosphere. The refracted waves now find a common focus with fast waves produced in other areas of the domain as they converge on the $a=c$ island surrounding the neutral point.  Meanwhile, the slow wave can easily be seen in parallel velocity images as it continues its passage guided by the field lines. 

The incoming fast waves spiral in to the upper area of the $a=c$ island, as further transmission/conversion occurs (see around $t=140$ s).  In contrast to the initial transmission/conversion observed around the lower $a=c$ layer, it is hard to delineate where small/large attack angles may occur.  This is due to the various trajectories of the incoming waves, the circular nature of the $a=c$ island, as well as the likelihood of multiple overlapping crossings of the $a=c$ layer.  It is safe to assume a wide range of attack angles are present, causing various amounts of transmission/conversion to occur.  These exchanges occur simultaneously with the slow wave arriving from underneath the neutral point. The ray description of the behaviour around the neutral point is discussed in Subsection \ref{D-X}.

Inside the island, $a<c$ and so the fast wave is again acoustically dominated. Hence converted fast (acoustic) waves propagate through this island at near the sound speed almost unhindered and travel essentially in straight lines. The transmitted slow magnetic waves, however, are highly bound to the fields lines and are seen to follow the separatrices emanating from the magnetic neutral point. Consequently, significant energy is funnelled both back towards the bottom boundary and out towards the top of the computational box.  Those waves propagating downwards will again meet the main $a=c$ layer and experience both transmission and conversion there. The last time slice depicted in Fig.~\ref{fig:vpar_linear} ($t=260.8$ s) shows that the beams persist below $a=c$, though they are no longer tied to the field lines. In fact, they refract to become perpendicular to the field, making clear they are fast (acoustic) beams in $a<c$.

Meanwhile, in the open field away from the neutral point, the slow waves continue to be guided by the field lines.  They freely propagate towards the top of the box and continue to grow in amplitude due to the ever decreasing density.

\subsubsection{Neutral Point Behaviour: Ray Description}  \label{D-X}

As discussed in Section \ref{S-X}, the behaviour of linear fast waves near the neutral point has been examined using a ray description in the low-$\beta$ limit by \cite{Nun17aa}, who found that ray paths describe logarithmic spirals and wave fronts (when concave) form caustics in a finite time that manifest as shocks. The introduction of a nonzero sound speed though introduces an $a=c$ equipartition circle about the point, near and inside which N\'u\~nez's analysis fails. In particular, the spiral will not complete.

\begin{figure}[htbp]
    \centering
    \includegraphics[width=\textwidth]{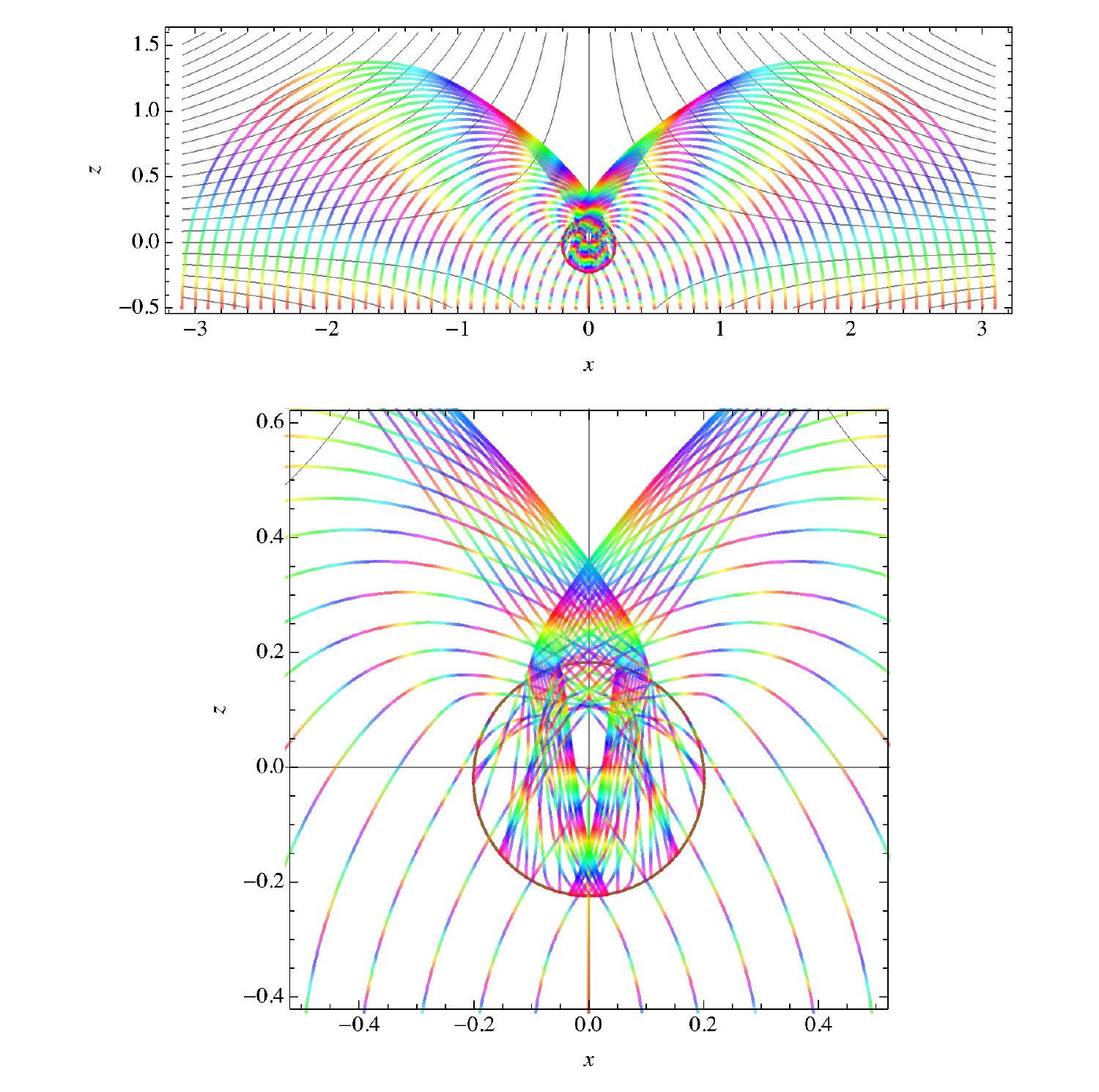}
    \caption{An ensemble of fast rays launched vertically from $z=-0.5$ attracted to the neutral point. The ray integrations are stopped upon leaving the island. No mode conversion is incorporated in the calculation, so the rays inside the island are pure fast (acoustic) rays and hence are nearly straight. Top: a full view of all rays; bottom: zoomed view of the neutral point and equipartition island. The colours used on the rays indicate phase, and make the concave structure of the wave fronts clear as they approach the neutral point. Caustics are particularly apparent along the top envelope and inside the island.}
    \label{fig:ray}
\end{figure}

Caustics are a well-known feature of optics and ray-tracing generally. They correspond to envelopes formed by rays \citep[][p.~247]{Whi74aa}, as seen for example along the top of the `arcades' shown in Fig.~\ref{fig:ray} (top panel) over $-1.4\lesssim x\lesssim 1.4$. Since the rays cross here, the energy density becomes infinite within the ray approximation, suggesting shock formation.\footnote{However, as pointed out by \cite{Whi74aa}, Sec.~8.8, this is an artefact of linearization. With shocks already present, the wave fronts speed up and typically push the rays apart, thereby inhibiting caustic formation.  Nevertheless, their appearance in linearized ray-tracing certainly suggests the development of nonlinearity in reality.}

In Figure \ref{fig:ray} we examine an archetypal isolated neutral point at the origin of the magnetic field $\B/\sqrt{\mu}=5(x,-z)$ with uniform sound speed $c=1$ and density $\rho(z)=e^{-z}$. These values prescribe the units of length, time and mass. The dispersion relation is taken simply as
\begin{equation}
   \mathcal{D}= \omega^4-(a^2+c^2)\omega^2 k^2+a^2 c^2 k^2 \kpar^2=0, \label{D}
\end{equation}
thereby neglecting the acoustic cutoff and {\bv} frequencies relative to wave frequency $\omega$.\footnote{Similar ray calculations were carried out by \cite{TarLin19aa} in 3D, including the Alfvén wave which we have neglected.} The wave vector $\k=(k_x,k_z)$ has absolute value $k$ and component in the field direction $\kpar=\k\thth\vdot\thth\widehat\B$. The ray equations \citep{Wei62aa} 
\begin{equation}
\frac{d {\bf x} }{ d\tau} = \frac{\partial\mathcal{D} }{ \partial {\bf k}}, 
 \quad
\frac{d {\bf {\bf k}} }{ d\tau} = -\frac{\partial\mathcal{D} }{ \partial {\bf x}},
\quad
\frac{dt }{ d \tau} = - \frac{\partial\mathcal{D} }{ \partial \omega},
\quad
\frac{d\omega }{ d \tau} = \frac{\partial\mathcal{D} }{ \partial t},
\quad
\frac{dS }{ d \tau} =\k\thth\vdot\thth\deriv{\x}{\tau},
\label{ray}
\end{equation}
are integrated numerically\footnote{A stiffness-switching scheme with projection is used to maintain $\mathcal{D}=0$ to high order.} from $z=-0.5$ and a range of $x$ values with starting wave numbers $k_x=0$ and $k_z$ calculated from $\mathcal{D}=0$ corresponding to an upward propagating fast wave. Keeping track of the phase $S$ allows us to identify wavefronts. 

Integrations are terminated on leaving the $a=c$ island. We do not take account of mode conversion, so the rays shown inside the island are pure fast (acoustic) rays, which are almost straight. In reality, they will be weakened and some energy transferred to near-field-aligned slow waves, though they are also partially focused, as seen in the lower panel.

The rays are coloured according to their phase, which is zero (red) initially in all cases. Since $\mathcal{D}$ is homogeneous in $\omega$ and $k$, the ray loci are independent of $\omega$, but the phase varies along them more quickly at higher frequencies. The illustrated case corresponds to $\omega=50$ in our dimensionless units.

It is striking that the bulk of the wave power is incident on the top of the equipartition island, where caustics have already formed, suggesting shock formation even for erstwhile-linear waves. This is entirely consistent with the linear simulation of Figure \ref{fig:vperp_linear}. The breadth of the launch region ($-3.1<x<3.1$ here) determines the compactness of the top ``hot spot''; a wider region broadens the focus too.

The structure of caustics is very dependent on initial conditions. Rays that are launched from below with some inclination from the vertical or from a greater depth may strike the $a=c$ island on the side rather than the top, though in all examined cases caustics typically form well before reaching it.

For the nonlinear simulation (see Sec.~\ref{S-shock-sim}), where the fast waves are already shocks well before encountering the neutral point environs, the shocks are further amplified by focusing and severely disrupt the $a=c$ island. 

The equipartition curve is notably the site of fast-to-slow mode transmission and conversion, so the oscillations inside the circle, where $c>a$, will be partially the transmitted slow (i.e., magnetically dominated near-transverse near-field-aligned) and partially the converted fast (acoustic) waves. The transmission and conversion coefficients $T$ and $C=1-T$ depend on attack angle \citep{SchCal06aa}, and so vary around the circumference of the circle.


Inside the island, the slow waves follow the field lines and their wavelengths decrease quadratically with distance from the neutral point, but approach it only asymptotically along the separatrices. On other (hyperbolic) field lines, they slowly pass by the neutral point before leaving the region through $a=c$ to again partially transmit and convert. The resulting slow waves in $a>c$ will depart the region, progressively more bound to field lines close to the separatrices. The resulting fast waves may again curl back towards the neutral point, and go through the process again. 

The fast (now acoustically dominated) waves inside the island travel near-isotropically along almost straight lines to depart the circle on the other side, where they too partially transmit (to slow/acoustic) and partially convert (to fast/magnetic) waves, the former depart and the latter may get caught up in the iterative process. A novel mathematical framework was developed by \cite{TraBriJoh12aa} for such an iterative scenario \citep[see also][Sec.~4.1.2]{TraBriRic14aa}, but the above description is sufficient for our purposes here.


Broadly then, the simple ray calculation, supplemented by a general understanding of fast/slow conversion, captures much of the behaviour of oscillations around the neutral point observed in our linear simulations and those of \cite{McLHoo06aa}, and informs our understanding of them. Specifically, (i) the reflecting fast waves converge from upper left and right on the neutral point; (ii) there is a significant increase in their amplitude as they do so (see Fig.~\ref{fig:vperp_linear}), and even the formation of a shock from a previously linear wave; (iii) complex behaviour takes place inside the island; and (iv) structured slow (field-parallel) waves emerge from the island along the separatrices (compare with the final two frames of Fig.~\ref{fig:vpar_linear}).

\subsection{Non-linear Shock Cases} 
  \label{S-shock-sim}
  
As in the linear case, a slab-like driver is implemented along the bottom boundary and is described by Equation (\ref{driver}).  The associated driver amplitude takes one of three values, $v_{0} = 1.5$ km s$^{-1}$, $v_{0} = 3.0$ km s$^{-1}$, or $v_{0} = 4.5$ km s$^{-1}$, which will be termed the weak, moderate and strong driver/shock cases respectively.  The weak driver amplitude is chosen such that the ensuing shock develops just before reaching the $a=c$ layer.  Accordingly, the moderate and strong drivers produce shocks that develop well before reaching the $a=c$ layer. Strong shocks are associated with greater compression ratios $X=\rho_2/\rho_1$, where $\rho_1$ and $\rho_2$ are the pre- and post-shock densities respectively, and also are faster.

\begin{figure}[thbp]
    \centering
    \includegraphics[width=\textwidth]{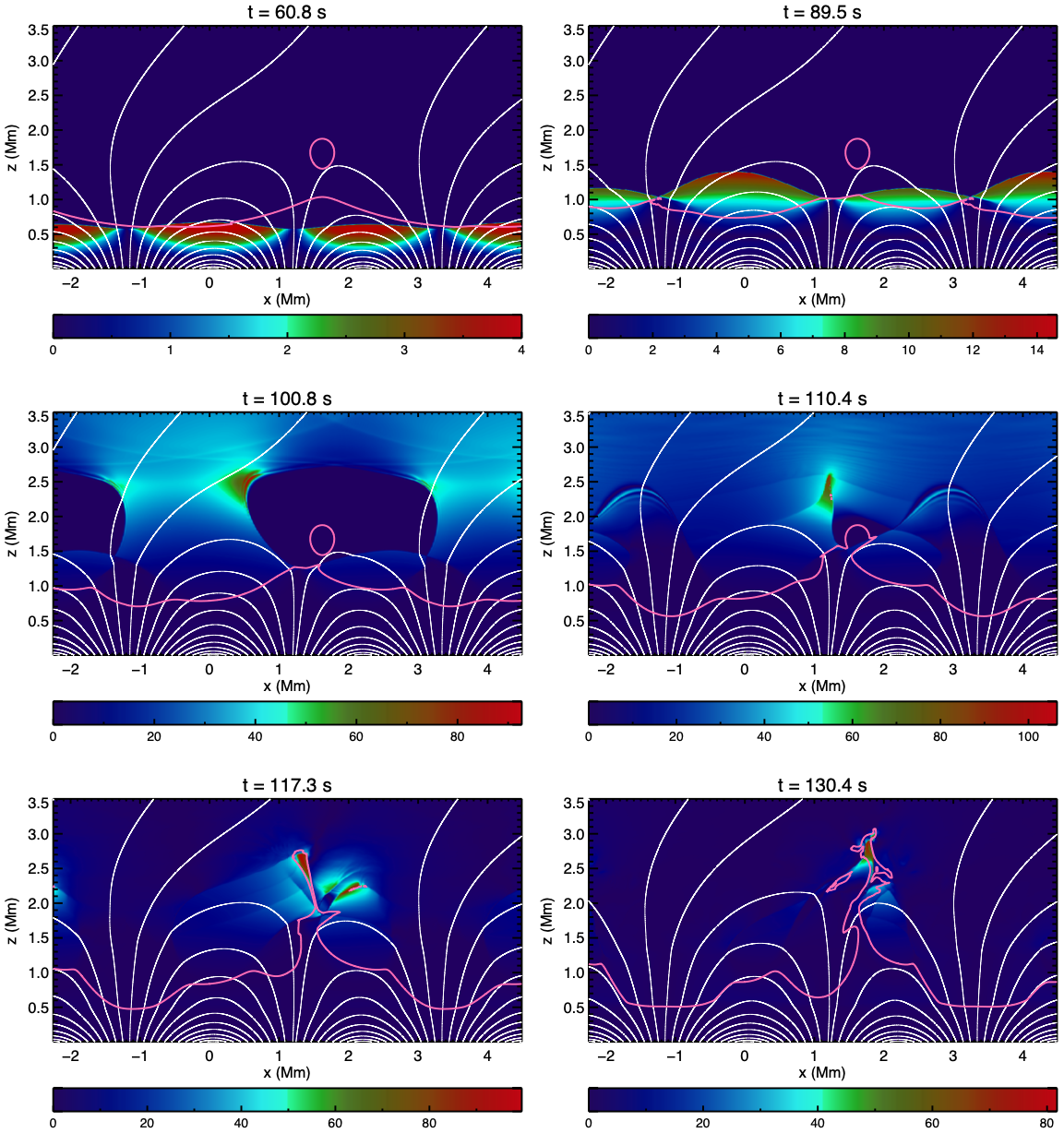}
    \caption{Timeline of $|v_{\perp}|$ snapshots for the weak shock case measured in km s$^{-1}$ with corresponding time in seconds displayed in each panel's title.  The superimposed white lines depict various magnetic field lines and the $a=c$ contours are represented in pink.}
    \label{fig:vperp_weak}
\end{figure}

\begin{figure}[thbp]
    \centering
    \includegraphics[width=\textwidth]{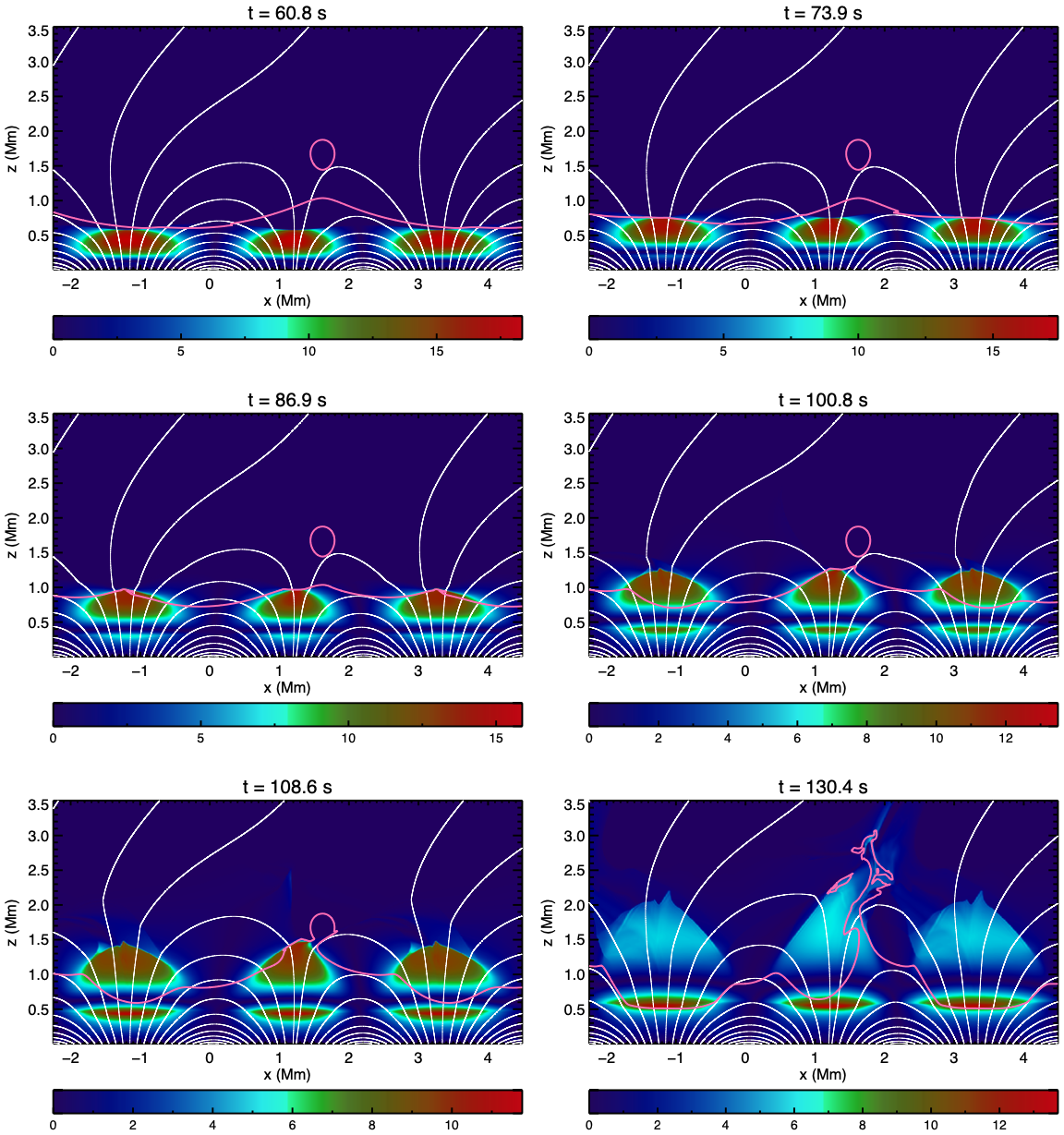}
    \caption{Timeline of $|v_{\parallel}|\sqrt{\rho}$ snapshots for the weak shock case measured in kg$^{1/2}$ m$^{-1/2}$ s$^{-1}$ with corresponding time in seconds displayed in each panel's title.  The superimposed white lines depict various magnetic field lines and the $a=c$ contours are represented in pink.}
    \label{fig:vpar_weak}
\end{figure}

\begin{sidewaysfigure}
    \centering
    \includegraphics[width=\textwidth]{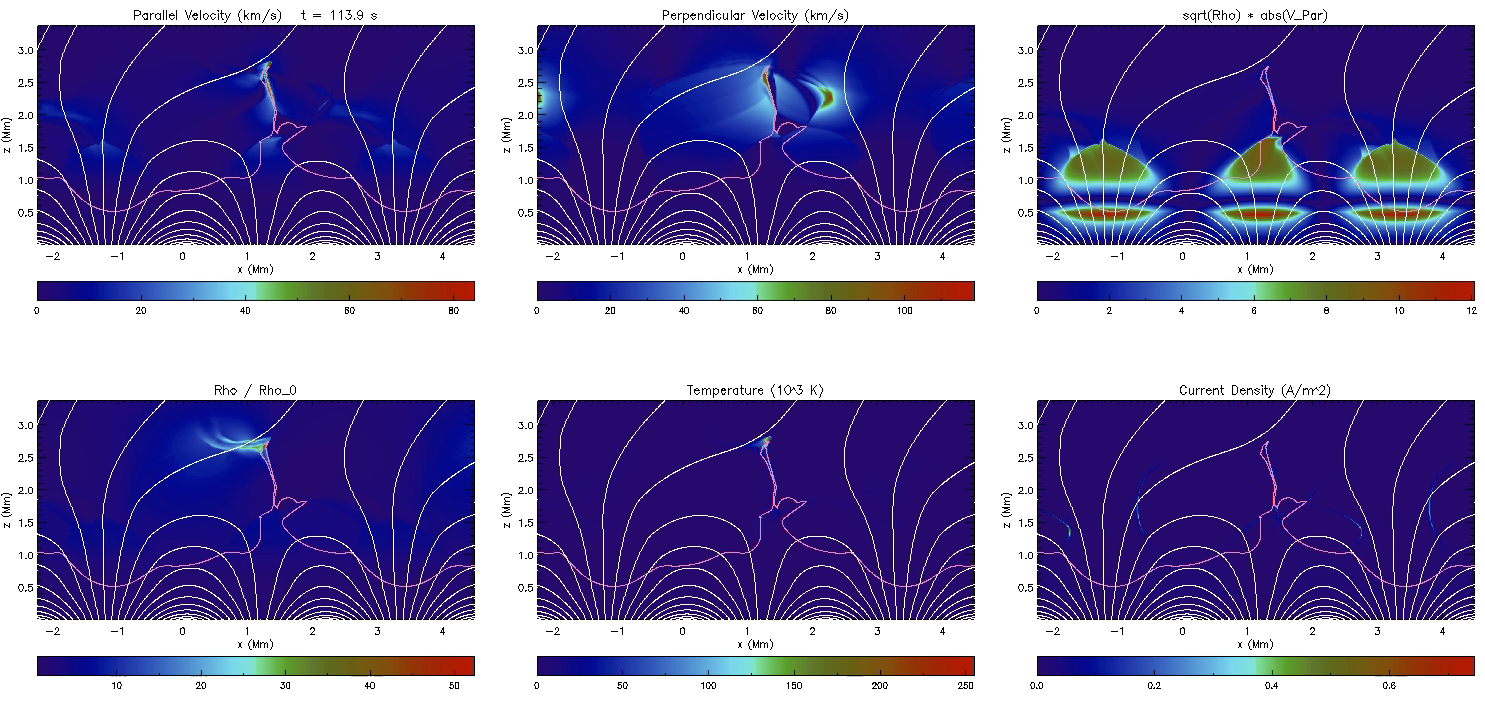}
    \caption{Snapshot at $t=113.9$ s from an animation (weak.mp4) of the weak shock case, showing respectively parallel velocity $v_\parallel$, perpendicular velocity $v_\perp$, scaled perpendicular velocity  $\sqrt{\rho}\,v_\perp$, density relative to the background state $\rho/\rho_0(z)$, temperature $T$ (in kK) and $y$ component of the current density, $j_y$. All quantities are shown as absolute values. The full animation is available as Electronic Supplementary Material.}
    \label{fig:weak_anim}
\end{sidewaysfigure}

A timeline of events in the weak shock case is displayed in Figures \ref{fig:vperp_weak} and \ref{fig:vpar_weak}, again showing both the perpendicular and (scaled) parallel plasma velocities respectively. The general description of fast/slow wave propagation is not dissimilar to that of the linear case described earlier, in particular the partial refraction of the fast wave towards the neutral point and the continued progress of the slow wave up the open field lines, as is evident in Fig.~\ref{fig:weak_anim} and its attached animation. However, the creation of non-linear shock waves produces some notable changes when compared to the linear case and these are described in detail in Section \ref{S-Discussion}.  The increasing driver strength between the shock simulations acts to enhance the phenomena identified and this is also discussed.

\begin{sidewaysfigure}
    \centering
    \includegraphics[width=\textwidth]{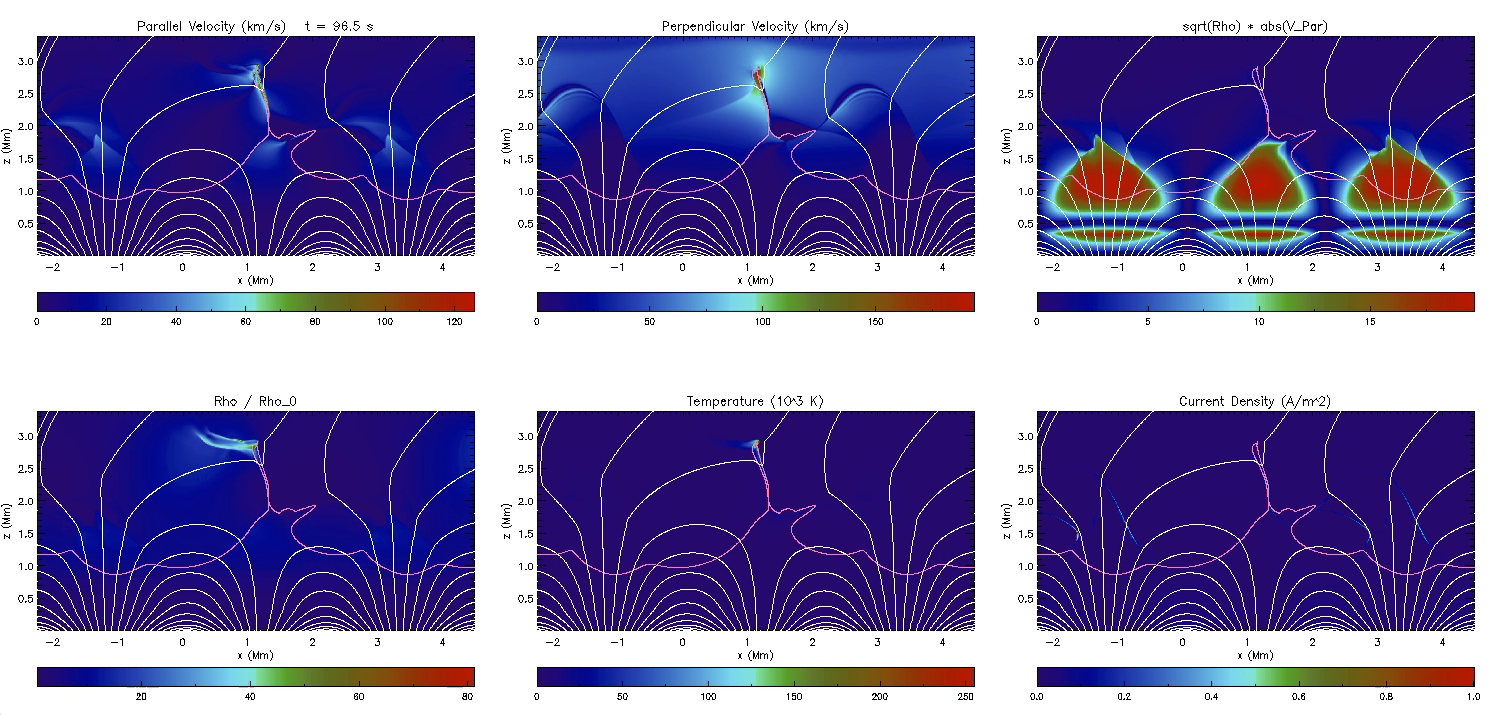}
    \caption{Snapshot at $t=96.5$ s from an animation (strong.mp4) of the strong shock case, showing respectively parallel velocity $v_\parallel$, perpendicular velocity $v_\perp$, scaled perpendicular velocity  $\sqrt{\rho}\,v_\perp$, density relative to the background state $\rho/\rho_0(z)$, temperature $T$ (in kK) and $y$ component of the current density, $j_y$. All quantities are shown as absolute values. The full animation is available as Electronic Supplementary Material.}
    \label{fig:strong_anim}
\end{sidewaysfigure}

Figures \ref{fig:weak_anim} and  \ref{fig:strong_anim} are frames from animations of the weak and strong shock cases respectively that show both parallel and perpendicular velocity components as well as density, temperature and current density. Unlike the linear case, the magnetic field lines and equipartition surfaces move substantially.

The strong shock (Fig.~\ref{fig:strong_anim} and animation) behaves similarly to the weak shock, though the $a=c$ equipartition level is dragged even further, and the magnetic island is disrupted even more violently. The behaviour of the different strength shocks is discussed in more detail in Section \ref{S-Discussion}.

\section{Analysis and Discussion} 
      \label{S-Discussion} 


In this article, we have sought to explore shock behaviour in a solar-like magnetic geometry exhibiting open and closed field and neutral points, to see how the basic behaviours of mode conversion, distinct fates of fast and slow waves, and shock smoothing translate and play out.

Animations corresponding to Figures \ref{fig:vperp_weak} and \ref{fig:vpar_weak} make it clear that the incident fast shock from below does indeed split into fast and slow components on reaching $a=c$. Furthermore, it is particularly clear in the $t=73.9$ s and $t=86.9$ s panels of Fig.~\ref{fig:vpar_weak} that the $a=c$ equipartition level in the open field region is dragged several hundred kilometres upward by the shock before bouncing back to near its original position.

\begin{figure}[htbp]
    \centering
    \includegraphics[width=\textwidth]{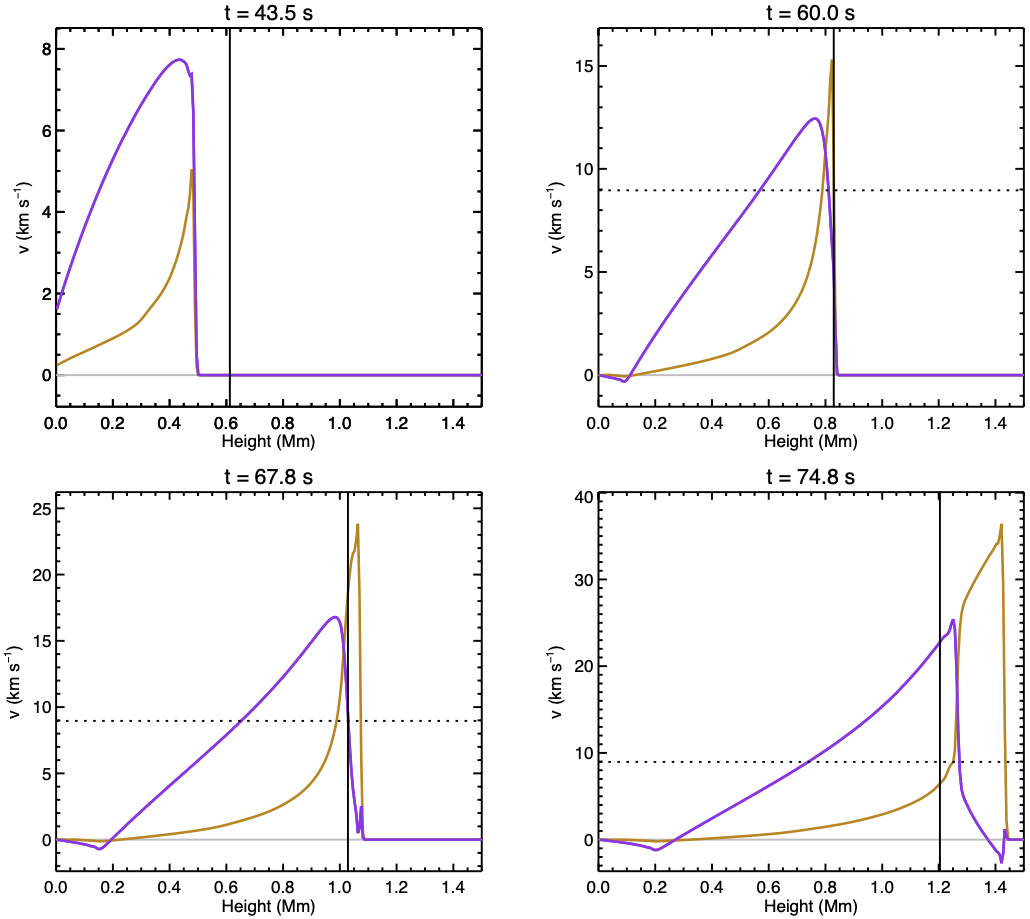}
    \caption{Velocity values taken along a vertical slice at $x = -1.05$ Mm for the strong shock case.  The purple and gold lines correspond to $v_{\parallel}$ and $v_{\perp}$ respectively, the dotted line is the equilibrium sound speed, and the black vertical line is where $a=c$.  The shock develops before reaching $a=c$ (top-left).  The slow shock (seen mainly in $v_{\parallel}$) is smoothed as it progresses through and drags the $a=c$ layer along with it (top-right and bottom-left) before steepening again to produce another shock front (bottom-right).  The fast shock (seen mainly in $v_{\perp}$) remains sharp for the duration and propagates unaltered.}
    \label{fig:smooth}
\end{figure}

This behaviour is also seen in Fig.~\ref{fig:smooth}, displaying parallel and perpendicular velocities in the strong shock along a vertical slice in the open field at $x=-1.05$ Mm. Here, the initial angle between the incoming shock front and the magnetic field varies from $12^{\circ}$ at $z=0.61$ Mm to $17^{\circ}$ at $z=1.20$ Mm -- these heights correspond to the initial and final location of the $a=c$ layer as the shock passes through. In the first panel ($t=43.5$ s), a single fast shock with both velocity components discontinuous approaches the $a=c$ level (vertical black line), reaching it in the second panel ($t=60$ s). By the third panel ($t=67.8$ s) it has dragged the $a=c$ layer over 200 km upward, and the perpendicular (gold) and parallel (purple) components have clearly separated into two distinct wave fronts, the former still discontinuous, but the latter now smoothed. By the final panel ($t=74.8$ s) the slow front has steepened again and a slow shock is reforming. The $a=c$ layer has been dragged a further 200 km by this stage. As the shock exits, the $a=c$ layer is seen to rebound back towards its initial equilibrium location (seen for the weak shock in Figs.~\ref{fig:vperp_weak} and \ref{fig:vpar_weak}, but more pronounced for the strong shock). This is consistent with the 1.5D results. In fact, the degree of acoustic smoothing seen in the second panel of Fig.~\ref{fig:smooth} is comparable to or in excess of that displayed in Fig.~2 ($\theta=15^\circ$) of \citet{PenCal19aa}. The small attack angle for the slice $x=-1.05$ Mm of Fig.~\ref{fig:smooth} results in clear but mild smoothing.

\begin{figure}[htbp]
    \centering
    \includegraphics[width=\textwidth]{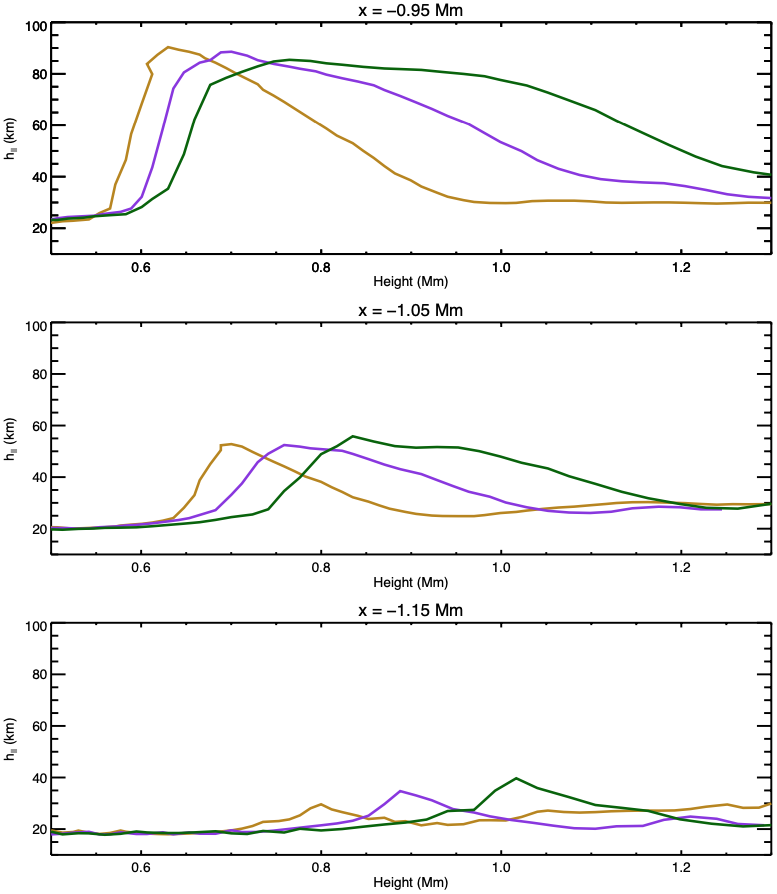}
    \caption{Velocity scale height $h_{\parallel}$ (see Equation (\ref{eq:hparallel})) at the steepest point of each individual shock front.  These values are taken along vertical slices at the $x$ coordinates given above each panel.  The gold, purple and green lines correspond to the weak, moderate and strong shocks respectively. The increased amplitude of the incoming shock results in greater dragging distances and hence larger regions over which the slow front is smoothed.}
    \label{fig:drag}
\end{figure}

To quantify the smoothing we define the velocity scale height
\begin{equation}
h_{\parallel}=\left|v_{\parallel,\text{max}}\middle/\left(\frac{dv_{\parallel}}{dz}\right)\right|
\label{eq:hparallel}
\end{equation}
as in \cite{PenCal19aa}, which provides a measure of the steepness of the parallel-velocity front. For a `mathematical' shock it should be zero, but for numerical reasons it does not drop much below about 20 km, even in a clearly fully developed shock.  In the absence of viscosity, the shortest scale-length will be the grid-resolution $\Delta z$ for this Lagrangian-remap scheme. However, shocks would lead to some Gibbs overshoot so shock viscosity is always applied, leading to an increased minimum scale length.  Values for $h_\parallel$ are depicted in Fig.~\ref{fig:drag} along three neighbouring vertical slices in the open-field region for weak, moderate and strong shocks. Here, it is plotted against height $z$ as the front propagated upwards, so greater $z$ corresponds to later times. 

Broadly, we see that the stronger shocks drag the $a=c$ layer further, and therefore that the smoothing is apparent at greater heights, both in terms of the initial smoothing event and the persistence. Specifically, at $x=-0.95$ Mm for example, the strong-shock front is still highly smoothed at $z=1.2$ Mm whereas the weak and moderate shocks have re-steepened considerably by this stage. Reference to Fig.~\ref{fig:vperp_weak} or \ref{fig:vpar_weak} shows that the magnetic field is highly inclined at $x=-0.95$ Mm, less so at $x=-1.05$ Mm and almost vertical at $x=-1.15$ Mm, thereby explaining the diminishing magnitude and persistence of the smoothing with this shift in $x$.


In the bottom left panel of Fig.~\ref{fig:vpar_weak}, we see that even for the weak shock the dragging of the $a=c$ layer causes a promontory to be created as the primary $a=c$ layer merges with that of the neutral point island. This is further deformed as the simulation progresses and as more waves converge on the area. The stronger the initial shock, the more deformation of the original $a=c$ equilibrium is observed. Because of the deformation of the $a=c$ island, the acoustic waves exiting this area are not as concentrated as they were in the linear case since they now have no common origin to be expelled from.

\begin{figure}[htbp]
    \centering
    \includegraphics[width=\textwidth]{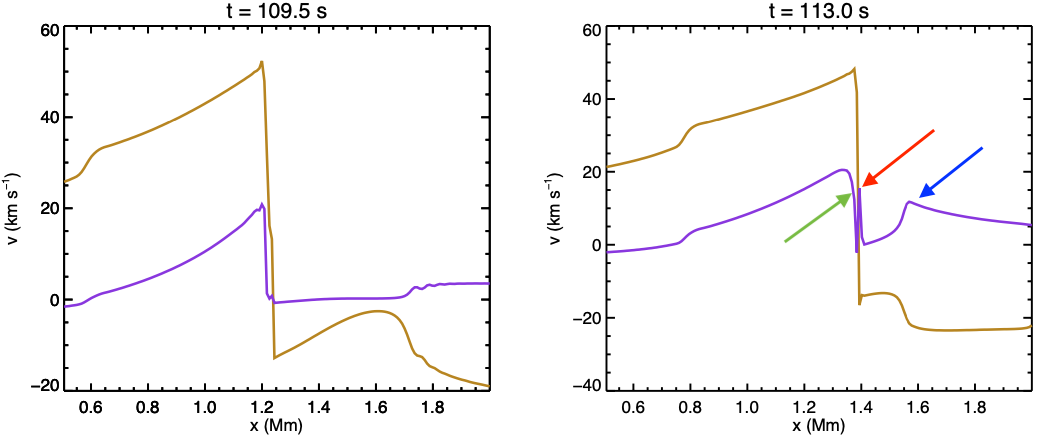}
    \caption{Velocity values taken along a horizontal slice at $z = 2.0$ Mm for the weak shock case at times depicted above each panel.  The purple line corresponds to $v_{\parallel}$ and the gold line to $v_{\perp}$. The central peaks correspond to a shock wave travelling to the right as it approaches the $a=c$ island/promontory from an area where $a>c$. As the wave reaches $a=c$, mode conversion occurs and smoothing is observed in the right-hand image for the slow component of $v_{\parallel}$. No appreciable smoothing is detected in $v_{\perp}$ or the fast component of $v_{\parallel}$. The red arrow indicates the fast shock, which is sharp in both parallel (purple) and perpendicular (gold) velocity components. The green arrow points to the smoothed slow wave, which shows predominantly in purple. The blue arrow indicates a disturbance impinging from the right.}
    \label{fig:island}
\end{figure}

This island/promontory presents the opportunity to explore a further question: does smoothing again occur when the incident fast shock propagates from $a>c$ to $a<c$? Figure \ref{fig:island} depicts the parallel and perpendicular velocities along a horizontal slice through the promontory just as the shock reaches $a=c$ (left) and a few seconds after (right), though it is still dragging the $a=c$ layer. The region at this stage is very dynamic, having already been impacted by a disturbance from the right, which is seen at $t=109.5$ s at about $x=1.75$ Mm and at $t=113$ s at around $x=1.65$ Mm. The splitting of the left-moving shock into fast and slow components is again apparent. At $t=113$ s, there is a clear fast front at about $x=1.4$ Mm, indicated by the red arrow, showing sharply in both velocity components. The slow shock is now most apparent in the parallel velocity, and is clearly smoothed (green arrow), though the fast shock also has a parallel component (the sharp purple spike). Shortly after the times depicted, the leftward and rightward waves collide, and interpretation becomes more difficult.


Finally, we turn to magnetic heating. The dissipation and attendant heating associated with currents in the solar atmosphere can take several forms. In the partially ionized quiet or active chromosphere, the ambipolar diffusion coefficient is often orders of magnitude larger than the ohmic diffusion term \citep[Figs.~5 to 7]{KhoColDia14aa}.
Both effects are enhanced by the production of small scales in turbulence, phase mixing, resonant absorption, etc., that locally amplify the current density $\bj=\curl\B/\mu$. Both shocks and neutral points generate small scales, and so are natural sites for electrical heating. 

\begin{figure}[thbp]
    \centering
    \includegraphics[width=.75\textwidth]{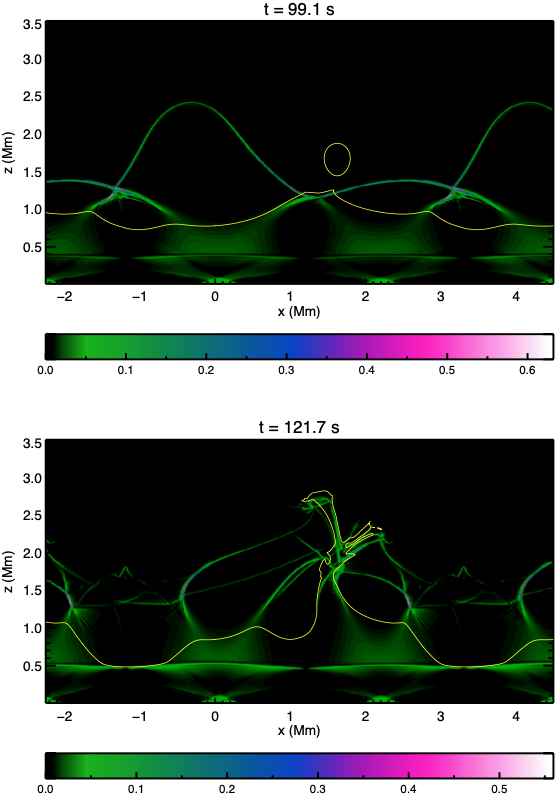}
    \caption{Absolute value of y-component of current density, $|\partial B_x/\partial z - \partial B_z/\partial x| / \mu$, in S.I units (A m$^{-2}$) for weak shock case ($v_{0}=1.5$ km s$^{-1}$) at times given above each panel, where the  $a=c$ layer is depicted in yellow.}
    \label{fig:curlb}
\end{figure}

Figure \ref{fig:curlb} depicts the $y$-component of the current density at two times for the weak shock. The top panel ($t=99.1$ s) is from shortly after the initial mode conversion at the primary equipartition level, and clearly shows the locations of both the fast and slow fronts. The neutral point island has not yet been disturbed. The magnitude of the peak current density is inevitably underestimated due to finite numerical resolution, but its sharpness is nevertheless made clear. The lower panel ($t=121.7$ s) shows how the disruption of the neutral point greatly enhances the opportunity for diffusive heating by creating much more fine structure, and this should be even more pronounced in 3D. Our simulations do not include 3D or diffusive terms, so these observations are qualitative, not quantitative, simply showing that transient sites favourable for dissipative heating are created by shocks, especially in concert with neutral points. However, in their MBP modelling, \cite{TarLin19aa} find that the attendant heating is still insufficient to supply radiative losses. Shock smoothing would further reduce heating.

\section{Conclusions}   \label{S-conclusions}
In summary, we have confirmed the shock smoothing process in a more realistic 2D model, and illustrated the differing paths and fates of the fast and slow shocks/fronts in open and closed field regions. Fast wave reflection away from high Alfv\'en speed regions is responsible for important solar processes even in the linear regime \citep[e.g., fast/Alfv\'en conversion and the generation of acoustic halos around active regions][]{CalHan11aa,RijRajPrz16aa}, but fast shocks have been shown to be well-capable of significantly disrupting neutral point structure and generating fine spatial scales. That fast shocks convert to fast shocks across equipartition surfaces is an important feature illustrated by our simulations.

It is also apparent that stronger shocks drag the equipartition layer further, and consequently the slow-wave smoothing interlude lasts longer in both time and space, which in some ways runs counter to intuition: the stronger the incident shock, the longer it takes the transmitted slow wave to re-shock.

Finally, we have verified for the first time that shocks moving from $a>c$ to $a<c$ split also (Fig.~\ref{fig:island}), and that the resulting slow component is again smoothed whilst the fast component remains sharp. This is consistent with the very small extent in $X$ of the slow lobe of the shock adiabatic seen in Fig.~7 of \cite{PenCal19aa}.







\bibliographystyle{spr-mp-sola}
\bibliography{fred}

\IfFileExists{\jobname.bbl}{} {\typeout{}
\typeout{****************************************************}
\typeout{****************************************************}
\typeout{** Please run "bibtex \jobname" to obtain} \typeout{**
the bibliography and then re-run LaTeX} \typeout{** twice to fix
the references !}
\typeout{****************************************************}
\typeout{****************************************************}
\typeout{}}

\end{article} 

\end{document}